\documentclass[%
 reprint,
 amsmath,amssymb,
 aps,
]{revtex4-1}

\usepackage{graphicx}% Include figure files
\usepackage{graphics}
\usepackage{dcolumn}% Align table columns on decimal point
\usepackage{bm}% bold math
\usepackage{float}
\usepackage{caption}
\usepackage{subfigure}
\usepackage{amsmath}
\usepackage{mathrsfs}
\usepackage{amssymb}
\begin{document}

\preprint{}

\title{Finite Size Effects of Thermal Conductivity for One-Dimensional Mesoscopic Systems}

\author{Li Wan}
\email{lwan@wzu.edu.cn}
\affiliation{Department of Physics, Wenzhou University, Wenzhou 325035, P. R. China}
\date{\today}% It is always \today, today,

\begin{abstract}
The finite size effects of the thermal conductivity $\kappa$ have been studied in the phonon space. It is found that only a few phonon modes are selected to take part in the thermal transport when the size $L$ of the system is decreased. The amount of the selected phonon modes is proportional to the $L$. In this way, $\kappa$ decreases with the decreasing of $L$. Such mechanism for the size effect of $\kappa$ found in this work is beyond the Phonon-Boundary scattering. The exponent $\alpha$ of the power law $\kappa \sim L^{\alpha}$ has been fitted, showing that the exponent is not universal.
\begin{description}
\item[PACS numbers]
44.10.+i,~  66.70.-f,~  65.80.-g, ~ 63.20.-e,~  63.22.-m.

\end{description}
\end{abstract}

%\pacs{66.70.-f,85.80.-g,65.40.gp,72.25.-b,63.21.-e,44.10.+i}% PACS, the Physics and Astronomy
\maketitle

%\tableofcontents
\section{Introduction}
With the development of nano technologies, electronic devices can be fabricated with scales down to mesoscopic size. In such scale, some of the physical properties of mesoscopic system change dramatically with the system size. For example, while thermal conductivity $\kappa$ for dielectric or semiconductor materials is a size independent in macroscopic system, it depends on the system size drastically in mesoscopic scale ~\cite{siemens,highland,sellan,koh,pop,Dhar,lepri}. The size effects on the $\kappa$ have been confirmed by the experiments~\cite{Wang,gang}. Especially the experiment basing on the transient thermal grating (TTG) shows that the $\kappa$ increases with increasing system size and is converged to a finite value for the infinite system size ~\cite{gang}. There exist two contradictions between such experimental results and the known theories. One is that calculations based on the harmonic interaction of atoms give rise to a divergent $\kappa$ even if the disorder and anharmonic interaction in the model are involved ~\cite{Dhar1,Gallavotti,Rieder,lepri1,Matsuda}. This is the so-called divergence problem in thermal transport. The second contradiction is the decreasing of $\kappa$ with the system size. Some authors have introduced a soft cut-off $l_c$ for the mean free path of phonons ~\cite{Maznev,gang}. They claimed that the phonons with the mean free path $l > l_c$ contribute little to the thermal transport. But how $l_c$ appear is still unknown. In this paper, we give a simple model which can resolve above two contradictions.\\

Generally, the thermal transport for one dimensional system is simplified as the coupling between oscillators ordered in chain. For a chain with a finite length, the oscillator number is noted by $L$ and the lattice parameter is $a$. The heat baths are attached to the two ends of the oscillator chain as a whole model~\cite{Dhar1,Gallavotti,Rieder}. It has been rigorously proved that if the coupling between the oscillators is harmonic, the thermal conductivity $\kappa$ diverges with the increasing of the size $La$ ~\cite{Rieder}. Even by invoking the disorder and anharmonic interaction in the model to enhance the thermal resistance, $\kappa$ is still not converged for the infinite size~\cite{lepri1,Matsuda}, which is the well-known divergence problem of $\kappa$~\cite{Dhar,lepri}. The divergence problem has been revealed by some other models in the real space, such as the hydrodynamics model~\cite{narayan}, and the model basing on the Mode-Coupling theory~\cite{lepri2}. According to those models, the relation between the $\kappa$ and $L$ follows a power law by $\kappa \sim L^{\alpha }$ with $\alpha >0$. The exponent $\alpha$ varies from model to model~\cite{Dhar,lepri}. \\

In the phonon space, the divergence problem is naturally removed by the phonon scattering. The phonon scattering enhances the thermal resistance and converges the $\kappa$ to be a finite value $\kappa_B \sim cv^2\tau$ for the infinite size ~\cite{peierls,ziman,Spohn,Pereverzev,Nickel,Maznev}. The phonon scattering has been implemented in the relaxation time $\tau$. Here, $c$ is the heat capacity of the system and $v$ is the averaged group velocity of phonon. The mechanisms for the phonon scattering are various, such as by Umklapp scattering ~\cite{ziman,Armstrong,Klemens,Klemens1}. Molecular Dynamics(MD) simulation has also been used to study the mechanism ~\cite{Alan,Sellan,Yang,Schelling}. In order to simplify the calculation, the approximation of single-mode relaxation time  $\tau _k$ has been applied to identify the contribution of each phonon at a wave vector $k$ in the Brillouin Zone(BZ). Then the $\kappa_B$ is modified to be $\kappa_B\sim \int c_k v^2_k \tau_k dk$ with the integral over the whole BZ ~\cite{peierls,ziman,Spohn,Pereverzev,Nickel,Maznev}. Unfortunately, no entire model covering the whole BZ is available for the $\tau_k$ yet. Some authors have proposed that $\tau_k \propto 1/\omega_k^2$ for phonon scattering at high frequency $\omega_k$ ~\cite{gang1,Klemens1,Mingo}. Such relation can not be extended to the low frequency range. Or, it is unphysical that $\tau_k$ goes to infinity when $\omega_k$ is approaching to the $\Gamma$ point of BZ, which also leads to an divergent $\kappa_B$. In order to simplify the phonon scattering for low frequency phonon, we make a assumption to modify the model by $\tau_k\propto 1/(\omega_k^2+C)$ for the extension. Here, $C$ a positive constant, dominating the $\tau_k$ when $\omega_k$ goes to zero. The introduction of $C$ into the $\tau_k$ is reasonable since the mean free path $v\tau$ of phonon has been measured to be a finite value ~\cite{Minnich}. \\

The size effect on the $\kappa$ can be obtained in the term of Phonon-Boundary(PB) scattering ~\cite{ziman,Sellan,Yang,Schelling,Maruyama}. The PB scattering means that phonons are scattered by the boundaries with the relation of $\tau \sim L$, which decreases the $\kappa$ for the finite system size ~\cite{Sellan}. But this is in fact incorrect in TTG case since the TTG is formed within a Si thin film and no any interfaces exist for the PB scattering ~\cite{gang}. In the TTG case, the authors have introduced a soft cut-off $l_c$ for the mean free path of phonons to show the size effect on the $\kappa$~\cite{Maznev,gang}. The $l_c$ means that the phonons with the mean free path $l>l_c$ contribute little to the thermal transport. But why does the TTG system have the $l_c$? In this work, we will show that only a few phonon modes are selected to take part in the thermal transport instead of the total phonons in the BZ. Such mechanism is beyond the PB scattering.\\

\section{Theory}
We consider the TTG as an uniformly ordered chain in one dimension with an infinite length. The oscillator number in one grating period is the $L$. The interaction between the oscillators is harmonic and the phonon scattering has been implemented in the relaxation time $\tau_k$. The phonon dispersion relation of the TTG can be obtained by using the lattice dynamics  theory and the wave vector $k$ is obtained by applying the Born-Von Karmen Boundary Condition(BKBC) on the chain ~\cite{ziman}. The total Hamiltonian $H$ of the system is the summation of the Hamiltonian of each oscillator by $H=\sum _m H_m$. Here, $H_m$ is the Hamiltonian of the $m^{th}$ oscillator, which is $H_m=p_m^2/(2M)+(1/2)\sum_n x_m \Phi_{mn}x_n$. In this expression, $x_m$ is the displacement of the $m^{th}$ oscillator deviating from its equilibrium position and $p_m$ is the momentum.  $\Phi_{mn}$ is the force constant for the interaction between the $m^{th}$ and the $n^{th}$ oscillators. And $M$ is the mass of one oscillator.\\

According to the Heisenberg's equation of motion, the time derivative of $H_m$ is $\dot{H}_m=\sum _n[\Phi_{mn}x_mp_n-\Phi_{nm}x_np_m]/(2M)$. On the right hand side of the equation, the first term  means the work done by all the other oscillators on the $m^{th}$ oscillator while the second term is for a reversed process. We introduce the creation operator $a^+_k$ and annihilation operator $a_k$ for phonon at $k$, and rewrite the total Hamiltonion as $H=\sum_k (a_k^+a_k+1/2)\hbar \omega_k$. The operators of $x_m$ and $p_m$ can be expressed as functions of $a_k$ and $a^+_k$, and then  the time derivative of $H_m$ reads
\begin{equation}
\label{eq1}
\dot{H}_m=\sum_{k,r}[e^{-i(k+r)R_m}T_{k,r}(t)+e^{i(k-r)R_m}W_{k,r}(t)]+h.c.
\end{equation}
Here, $T_{k,r}$ and $W_{k,r}$ are time-dependent operators, as functions of $a_{k(r)}$ and $a^+_{k(r)}$ in the phonon space. $R_m$ is the equilibrium position of the $m^{th}$ oscillator.\\

Since the energy of each oscillator is real and periodical in the chain by $H_{m+L}=H_m$, it can be obtained from the eq.(\ref{eq1}) that $k$ or $r$ takes only the values of 
\begin{equation}
\label{eq2}
\Lambda=\frac{2\pi h}{La}, ~~~~h=0,\pm 1,\pm 2,\cdots, \pm \frac{L}{2}.
\end{equation}
That means only the phonon at $\Lambda$ are selected to take part in the time evolving of $H_m$ and are responsible for the decaying of the TTG. All the other phonon modes not at $\Lambda$ in the BZ act as a constant thermal background, and are eliminated after the time derivative. For the selected phonon, the nearest wave vector close to the $\Gamma$ point is $k_c=\pm 2\pi /(La)$ with $h=\pm 1$, indicating that phonon in the range of $0<|k|<|k_c|$ contributes little to the thermal transport even though with their large mean free paths. Thus, $k_c$ acts as the origin for the soft cut-off $l_c$ proposed in the models for the TTG~\cite{gang,Maznev}. The amount of the selected phonon modes $\Lambda$ equals $L$. That means fewer phonon modes are selected for the thermal transport for smaller $L$ by modulating the thermal grating. In this way, the size effect of the $\kappa$ can be clarified. This issue will be specified later. \\

Before we calculate the $\kappa$ for the TTG, we turn to another type of the oscillator chains with a finite length of the total chain, comparing to the TTG with a finite length of one period but an infinite length of the total chain. Such finite-size chain(FSC) can be found like the nanotubes ~\cite{Wang}. We will get the similar result of the TTG for the FSC. The FSC is a uniformly ordered chain in one dimension with two open ends, and $L$ now represents the total number of the oscillators of the chain. For convenience, we index the ordered oscillators from $0$ to $L-1$. The Hamiltonian of the chain reads
\begin{equation}
\label{eq3}
H_f=\sum _{m}\frac{M\dot{x}_m \dot{x}_m}{2}\Theta_m+\frac{1}{2}\sum _{m,n}x_m \Phi_{m,n}x_n\Theta_m\Theta_n.
\end{equation}
Here, $\Theta_{m(n)}$ is the geometric structure factor(GSF) of the chain, and defined as $\Theta_{m(n)} =1$ for $0\le m(n)\le L-1$ and $\Theta_{m(n)}=0$ for other cases. By using the GSF, the FSC is extended to be infinitely long without any changing to the total Hamiltonian. The motivation for such extension is that the lattice dynamics theory and the BKBC can only be applicable on the chain with an infinite length. In this way, we can get the basic phonon dispersion relation as a start point for the study of FSC. We use $N$ to denote the total number of oscillators of the extended chain with $N\rightarrow \infty$. We rewrite the Hamiltonian by using the Fourier transformation $x_m=(1/\sqrt{NM})\sum _k e^{i k R_m}Q_k$. The $Q_k$ is the normal coordinate of phonon at $k $ for the extended chain. The Hamiltonian then reads
\begin{equation}
\label{eq4}
H_f=\frac{1}{2}\sum_{k,r}\dot{Q}_k\dot{Q}_r^*Z_{k-r}+\frac{1}{2}\sum_{k,r,q}\omega_q^2Z_{q-k}Z_{r-q}Q_k^*Q_r.
\end{equation}
Here, $Z_k$ is the GSF defined in the phonon space, which is obtained by the Fourier transformation of $\Theta_m$. The $Z_k$ has an expression of $Z_{k}=\frac{e^{ikLa}-1}{N(e^{ika}-1)}$. The basic phonon dispersion relation  $\omega_q\sim q$  of the extended chain is obtained by the Fourier transformation of the force constant  $\Phi_{mn}$. It can be found that the normal coordinates $Q$ are coupled to each other due to the finite size effect. It is also found that when $L$ is approaching to $N$, the GSF $Z_k$ goes to $\delta_{k,0}$ and eq.(\ref{eq3}) recovers the diagonalized Hamiltonian $H_f=\frac{1}{2}\sum_{k}\dot{Q}_k\dot{Q}_k^*+\frac{1}{2}\sum_{k}\omega_k^2Q_k^*Q_k$ for a real infinitely long chain.\\

We introduce a new set of coordinates $\Xi$ in the phonon space by the definition of $\Xi_p=\sum_r Q_rZ_{r-p}$. The motion equation of the $\Xi_p$ then reads
\begin{equation}
\label{eq5}
\frac{\partial ^2 \Xi_p}{\partial t^2}+\frac{a}{2\pi}\int_{-\pi/a}^{\pi/a}\omega_k^2 \frac{e^{iLa(k-p)}-1}{e^{ia(k-p)}-1}\Xi_kdk=0
\end{equation}
in integral form. The rigorous numerical solution to the integral equation is not applicable due to the highly oscillated integral kernel. It is expected that the density distribution of the phonon eigenmodes as the solutions are narrow peaks located at some wave vectors. We approximate the density peaks as $\delta$ functions and denote the locations of the eigenmodes by $p$ or $q$ in the following. Thus, the density of states (DOS) of $p$ mode phonon is noted by $2\pi \eta_p \delta(k-p)$. The DOS is used to transform the eq.(\ref{eq4}) to be in the form of summation, reading 
\begin{equation}
\label{eq6}
\Omega_p^2 \Xi_p=a\sum_q\omega_q^2 \frac{e^{iLa(q-p)}-1}{e^{ia(q-p)}-1}\Xi_q \eta_q
\end{equation}
with $\Omega_p$ the frequency of the $p$ mode and $\Xi_p\sim e^{-i \Omega_p t}$. The phonon eignmodes of the solution should be decoupled to each other, meaning $q-p=2\pi h/(La)$ must be held to reduce the eq.(\ref{eq6}) to be $\Omega_p^2 \Xi_p=aL \eta_p\omega_p^2 \Xi_p $. Here, $h$ represents integers. Then, one obtains that the locations of the phonon eigenmodes $p(q)$ of the FSC are at
$\Lambda=2\pi h/(La)$ with $h$ the integers in the range of $[-L/2,L/2]$. This result is the same as that of TTG, only different in meaning of $L$. When $L$ is approaching to $N$ for the infinitely long chain, the equation of motion for the $p$ mode should recover $\ddot{\Xi}_p=-\omega_p^2 \Xi_p$. That means the eigenfrequency $\Omega_p$ must be equal to $\omega_p$ and $\eta_p$ equals $1 /(La)$. Physically, $2\pi\eta_p$ is the degeneracy of the $p$ mode per unit length of the FSC. It is interesting that applying the BKBC directly on the FSC by using the plane waves as the lattice waves can get the same result of $\Lambda$. However, it should be emphasized that the direct application of the BKBC is not rigorous for the study, since the DOS of phonon eignmodes are not exactly the $\delta$ functions according to eq.(\ref{eq5}).\\

Fig.1(a) shows the phonon dispersion relation of the FSC by the discrete data points while the solid continuous line is the relation for the infinitely long chain. For the calculation in this paper, the dispersion relation of $\omega(k)=\omega_0 |\sin(k/2)|$ is used as an example with $\omega_0=4\times 10^{12}Hz$ of Silicon applied ~\cite{zhang1}. And the lattice parameter is set to be $a=1$ for simplicity. 
\begin{figure}
\subfigure{
\label{fig1a}
\includegraphics[width=6cm]{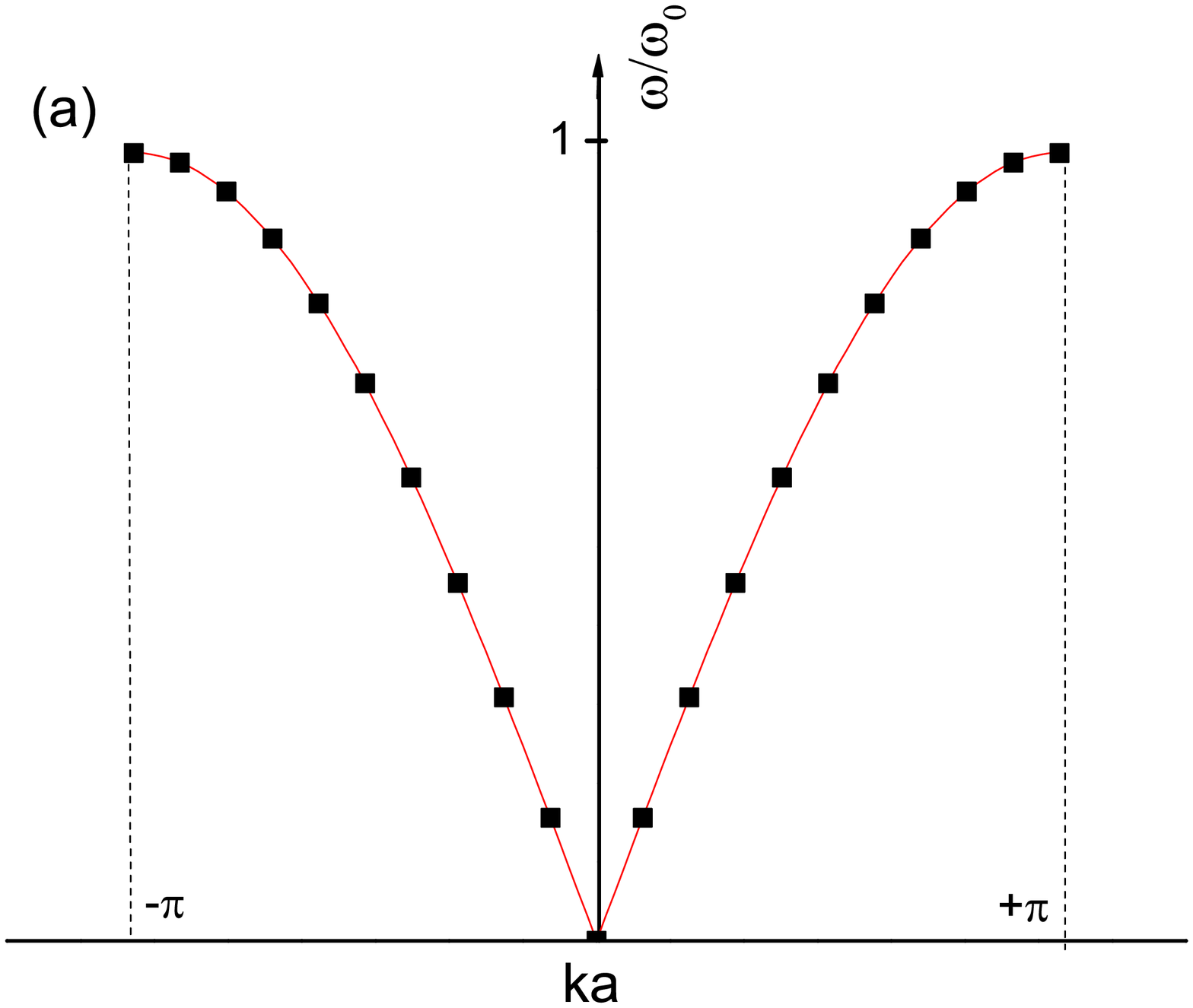}}\\
\subfigure{
\label{fig1b}
\includegraphics[width=6cm]{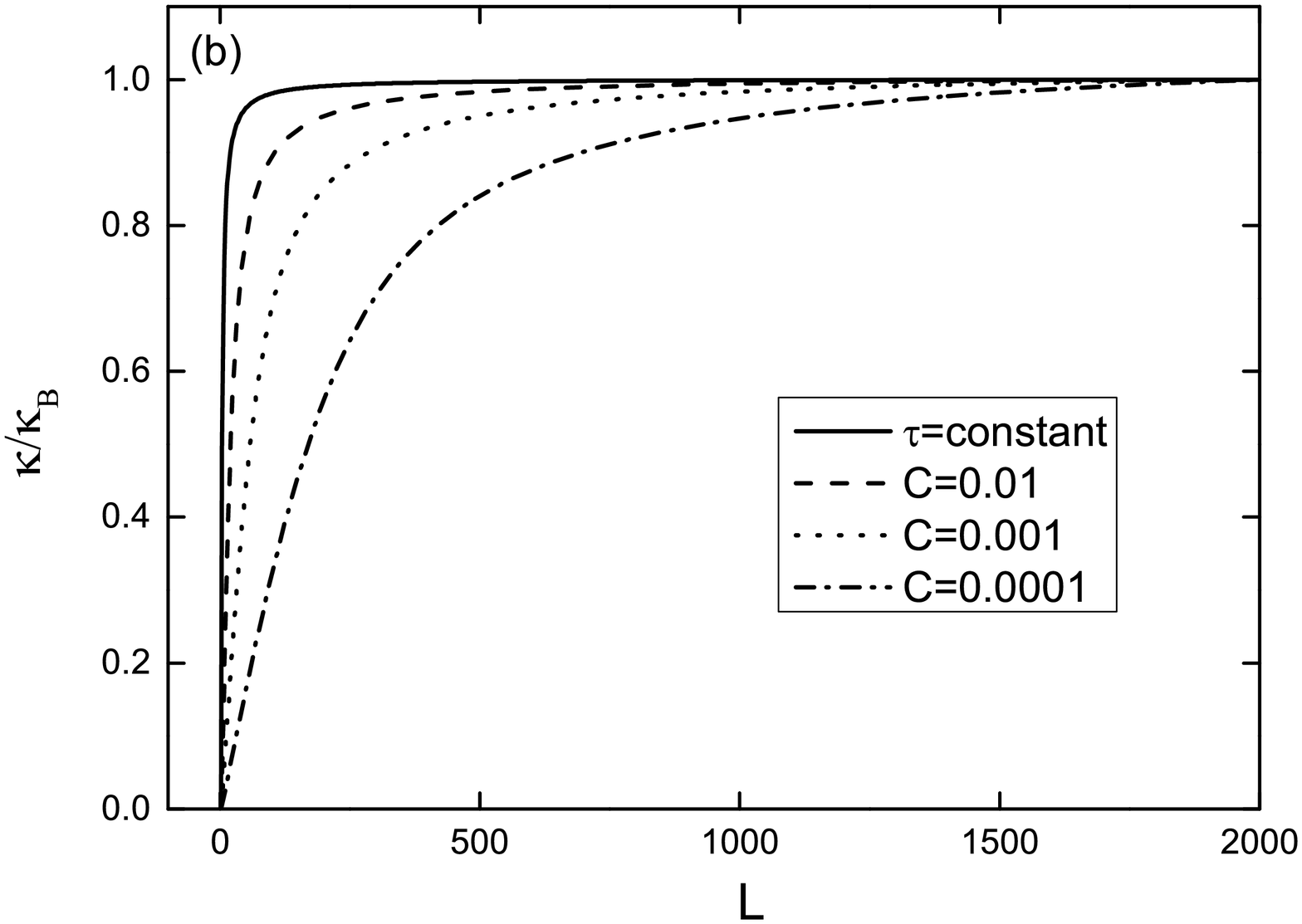}}\\
\subfigure{
\label{fig1c}
\includegraphics[width=6cm]{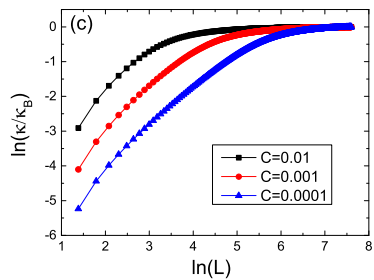}}
\caption{(a) Phonon Dispersion Relations(PDR). Solid and continuous line is the PDR for the infinitely long chain while the discrete data points are for the chain with a finite size. In the plot, $L=20$ is used. (b)The renormalized thermal conductivity $\kappa/\kappa_B$ as a function of the system size $L$ with various $C$. In the plot, $C$ have been renormalized by $\omega_0^2$. (c) $\ln(\kappa/\kappa_B)$ as a function of $\ln(L)$, showing the power law of $\kappa \propto L^{\alpha}$ for small $L$. The exponent $\alpha$ can be fitted from the curves.}
\end{figure}
 Comparing to the solid line, phonon modes not at $\Lambda$ are absent for the FSC. Especially, the phonon in the range of $-2\pi/(La)<k<2\pi/(La)$ have no contribution to the thermal transport for the FSC. The absence of the phonon close to the $\Gamma$ point for the thermal transport has been confirmed by the TTG experiment~\cite{gang}.\\ 
 
\section{Results and Discussion}
We have studied the phonon modes of two types of the mesoscopic structures. One type of the structures is the TTG and the other type is the FSC. The two types of the structures have the same result that only a few phonons located at $\Lambda$ can take part in the thermal transport. We call these phonons by selected phonons. The total amount of the modes of the selected phonons equals $L$. The period length of TTG or the total length of the FSC, both of which have been denoted by $L a$ in this paper, can be changed to bring three effects on the phonon behaviors. Suppose the $L$ is decreased as an example to specify the three effects. The first effect is that the mode amount $L$ of the selected phonons then is reduced in the BZ. The second effect is that the $\eta=1/(La)$ is enhanced for each selected phonon mode. Actually, these two effects are canceled for the thermal transport due to the produce of the mode amount and the degeneracy as a constant of $L\eta=1/a$. The third effect is that the selected phonon mode close to the $\Gamma$ point are shifted to higher frequencies by $\omega=\omega_0\sin(\pi/(La))$. According to the Bose-Einstein distribution $f_k(T)=(e^{\hbar \omega_k/(k_B T)}-1)^{-1}$, higher frequency state has fewer averaged phonon amount at a fixed temperature. Thus, the third effect leads to fewer phonons taking part into the thermal transport, which is the key mechanism to the size effect on $\kappa$.\\

Now we jump to the calculation of $\kappa$. The general expression of the $\kappa$ reads
\begin{equation}
\label{eq7}
\kappa = 2\pi\sum_{p=\Lambda} \eta_p c_p v_p^2 \tau_p.
\end{equation} 
Here, $2\pi\eta_p$ has been understood as the degeneracy of the selected phonon mode. For the FSC, we have obtained $\eta_p=1/(La)$. And for the TTG, $\eta_p$ is a product of two factors by $\eta_p=\eta_1 \cdot \eta_2$. Here, $\eta_1=1/(Na)$ is contributed from the infinite length $Na$ of the TTG, while $\eta_2=N/L$ is the amount of the grating periods as copies of the thermal transport in the real space. Thus, $\eta_p=1/(La)$ is the same for the TTG. And, $c_p$ is the heat capacity for the $p$ phonon mode while  $ \eta_p c_p$ is the specific heat due to the $\eta$ inversely proportional to $La$. The $c_p$ has the expression of $\hbar \omega_p \partial f_k(T)/\partial T$. The phonon group velocity $v_p$ has the definition of $v_p=\partial \omega(k)/\partial k|_{k=p}$. The relaxation time $\tau_p$ is still within the single-mode approximation, and has been modified to be $\tau_k\propto 1/(\omega_k^2+C)$ as discussed above. In the calculation, the summation is over the discrete modes $\Lambda$ as obtained in eq.(\ref{eq2}). By setting $L=N$, the $\eta_p$ then acts as the infinitesimal in the BZ, transforming the eq.(\ref{eq7}) in integral form to recover $\kappa_B \sim \int c_kv_k^2\tau_k dk$.\\

\begin{table}[tbp]
\caption{The exponent $\alpha$ of the power law  $\kappa\sim L^{\alpha}$ fitted from the curves in fig1.(c) with various $C$. The error for the fitting has been indicated. Results show that the exponent $\alpha$ is not universal.}
\begin{tabular}{c|c|c|c}
\hline
~~~~C~~~~ & ~~~~0.01~~~~ & ~~~~0.001~~~~ & ~~~0.0001~~~ \\
\hline
~$\alpha$~ & 1.167 & 1.324 & 1.34 \\
\hline
~error($\pm$)~ & 0.0486 & 0.03247 & 0.03017 \\
\hline
\end{tabular}
\end{table}

Fig.1(b) is the plot of $\kappa/\kappa_B$ as a function of $L$ with various $C$. In the plot, $C$ has been renormalized by $\omega_0^2$ for clarity. The plot indicates the size effects on the $\kappa$, which matches the behaviors of $\kappa$ measured by experiments ~\cite{Wang,gang}. The $\kappa$ converges to $\kappa_B$ for $L \rightarrow \infty$. The mechanism of the selected phonons is beyond the Phonon-Boundary scattering, and can be used to understand the size effect on $\kappa$. In order to study the exponent $\alpha$ for the power law  $\kappa\sim L^{\alpha}$, we have plotted the $\ln (\kappa/\kappa_B)$ as a function of $\ln (L)$ in fig.1(c). The exponents $\alpha$ are fitted from the curves in the range of $6<L<20$, and the results are listed in table 1. It shows that $\alpha$ is not universal, and dependents on the detail phonon scattering from material to material. \\[0.1cm]

\section{Conclusion}
As a conclusion, we have studied the finite size effect on the thermal conductivity $\kappa$ in the phonon space. The $\kappa$ decreases with the decreasing of the size of the structures. The selected phonon modes have been obtained to interpret the size effect, which is beyond the Phonon-Boundary scattering. The exponents $\alpha$ of the power law $\kappa \sim L^{\alpha}$ have been fitted, showing that the exponent is not universal.\\
%\nocite{*}
\appendix 
\section{}
In this appendix, we complete the derivation of eq.(1) in the main text. The Hamiltonian of the $m^{th}$ oscillator reads
\begin{equation}
\label{eq8}
H_m=\frac{p_mp_m}{2M}+\frac{1}{2}\sum \limits_n x_m \Phi_{ms}x_n.
\end{equation}
According to Heisenberg equation, one obtains the following equation after some algebra
\begin{equation}
\label{eq9}
\dot{H}_m=\frac{i}{\hbar}[H,H_m]=\sum _n \frac{1}{2M}[\Phi_{m,n}x_mp_n-\Phi_{n,m}x_np_m].
\end{equation}
 We define the Fourier transformations
\begin{equation}
\label{eq91}
\begin{split}
&x_m=\frac{1}{\sqrt{NM}}\sum \limits _{k} e^{ikR_m}Q_{k};~~~Q_k=\sqrt{\frac{M}{N}}\sum \limits _{m} e^{-ikR_m}x_m,\\
&p_m=\sqrt{\frac{M}{N}}\sum \limits _{k} e^{-ikR_m}P_{k};~~~P_k=\sqrt{\frac{1}{MN}}\sum \limits _{m} e^{ikR_m}p_m,
\end{split}
\end{equation}
with $Q$ and $P$ the coordinates and momenta in the phonon space. $R_m$ is the equilibrium position of the $m^{th}$ oscillator. Since the $x$ and $p$ are real, one gets $Q_k^*=Q_{-k}$ and $P_k^*=P_{-k}$.
By using the create and annihilation operators 
\begin{equation}
\begin{split}
\label{eq92}
&a_k=\sqrt{\frac{\omega_k}{2\hbar}}\left (Q_k-\frac{P_k^*}{i\omega_k}\right ),\\
&a^+_k=\sqrt{\frac{\omega_k}{2\hbar}}\left (Q_k^*+\frac{P_k}{i\omega_k}\right ),
\end{split}
\end{equation}
we have 
\begin{equation}
\label{eq10}
Q_k=\sqrt{\frac{\hbar}{2\omega_k}}(a_k+a^+_{-k});~~~~P_k=i\sqrt{\frac{\hbar \omega_k}{2}}(a_k^+-a_{-k}),
\end{equation}
Substituting eq.(\ref{eq10}) into the eq.(\ref{eq91}), one obtains
\begin{equation}
\label{eq11}
x_m=\sum \limits _{k}\sqrt{\frac{\hbar}{2NM\omega_k}} [e^{i(kR_m-\omega_k t)}a_k(0)+e^{-i(kR_m-\omega_k t)}a_k^+(0)],
\end{equation}
\begin{equation}
p_m=\sum \limits _{k} i\sqrt{\frac{\hbar M\omega_k}{2N}}[e^{-i(kR_m-\omega_k t)}a_k^+(0)-e^{i(kR_m-\omega_k t)}a_k(0)].
\end{equation}
Then the motion equation eq.(\ref{eq9}) reads
\begin{equation}
\label{eq12}
\dot{H}_m=\sum_{k,r}[e^{-i(k+r)R_m}T_{k,r}(t)+e^{i(k-r)R_m}W_{k,r}(t)]+h.c,
\end{equation}
with
\begin{equation}
\label{eq13}
T_{k,r}(t)=\frac{\hbar i}{4MN}\sqrt{\frac{ \omega_r}{\omega_k}}[\omega_r^2-\omega_k^2]e^{i(\omega_r t+\omega_k t)}a_k^+(0)a_r^+(0);
\end{equation}
\begin{equation}
\label{eq14}
W_{k,r}(t)=\frac{\hbar i}{4MN}\sqrt{\frac{ \omega_r}{\omega_k}}[\omega_r^2-\omega_k^2] e^{i(\omega_r t-\omega_k t)}a_k(0)a_r^+(0).
\end{equation}

\section{}
In this appendix, we complete the derivations of eq.(\ref{eq4}) and eq.(\ref{eq5}) in the main text. The ordered oscillators of the finite chain are indexed from $0$ to $L-1$. The Hamiltonian of the FSC in the main text is the  eq.(\ref{eq3}), repeated here by
\begin{equation}
\label{eq15}
H_f=\sum _{m}\frac{M\dot{x}_m \dot{x}_m}{2}\Theta_m+\frac{1}{2}\sum _{m,n}x_m \Phi_{m,n}x_n\Theta_m\Theta_n.
\end{equation}
By using the Fourier transformation $x_m=\frac{1}{\sqrt{NM}}\sum _k e^{i k R_m}Q_k$, one rewrites the Hamiltonian  as
\begin{equation}
\label{eq16}
\begin{split}
H_f=&\sum_{m,k,r}\frac{\Theta_m}{2N}e^{iR_m(r+k)}\dot{Q}_k\dot{Q}_r \\
&+\sum_{m,n,k,r}\frac{\Phi_{m,n}\Theta_m\Omega_n}{2NM}e^{i(rR_n+kR_m)}Q_kQ_r.
\end{split}
\end{equation}
Here, $N$ is the total number of the oscillators of the extended chain with $N \rightarrow \infty$, which has been specified in the main text. For an infinitely long chain, there exists a relation between the force constant $\Phi$ and the phonon frequency $\omega_k$ by
\begin{equation}
\label{eq17}
\begin{split}
&\frac{1}{\sqrt{N}}\sum _m e^{-i k(R_m-R_n)}\Phi_{mn}=\frac{M}{\sqrt{N}}\omega_k^2, \\
&\Phi_{mn}=\frac{1}{N}\sum_k M\omega_k^2 e^{ik(R_m-R_n)}.
\end{split}
\end{equation}
We define $Z_{k}$ by
\begin{equation}
\label{eq18}
Z_{k}=\sum_m\frac{\Theta_m}{N}e^{iR_mk}=\frac{1}{N}\frac{e^{ikLa}-1}{e^{ika}-1}.
\end{equation}
Actually, $Z_{k}$ is the geometric structure factor in the phonon space. The Hamiltonian then reads
\begin{equation}
\label{eq19}
\begin{split}
H_f&=\sum_{m,k,r}\frac{\Theta_m}{2N}e^{iR_m(r+k)}\dot{Q}_k\dot{Q}_r  \\
&+\sum_{m,n,k,r}\frac{\Phi_{m,n}\Theta_m\Omega_n}{2NM}e^{i(rR_n+kR_m)}Q_kQ_r \\
&=\frac{1}{2}[\sum_{k,r}Z_{r-k}\dot{Q}_k^*\dot{Q}_r+\sum_{k,r,q}\omega_q^2 Z_{q-k}Z_{r-q}Q_k^*Q_r]. 
\end{split}
\end{equation}
This  is the eq.(\ref{eq4}) in the main text. The Lagrange function $\mathcal{L}$ is 
\begin{equation}
\label{eq20}
\mathcal{L}=\frac{1}{2}[\sum_{k,r}Z_{r-k}\dot{Q}_k^*\dot{Q}_r-\sum_{k,r,q}\omega_q^2 Z_{q-k}Z_{r-q}Q_k^*Q_r].
\end{equation}
The conjugate momentum $P_q$ of the $Q_q^*$ in the phonon space can be obtained as
\begin{equation}
\label{eq21}
P_q=\frac{\partial \mathcal{L}}{\partial \dot{Q}_q^*}=\frac{1}{2}\sum_k(\dot{Q}_kZ_{k-q}+\dot{Q}_k^*Z_{-q-k})=\sum_k\dot{Q}_kZ_{k-q}.
\end{equation}
And then the equation of  motion reads
\begin{equation}
\label{eq22}
\begin{split}
\dot{P}_q=-\frac{\partial H}{\partial Q_q^*}=-\sum_{k,r}\omega_k^2Q_rZ_{k-q}Z_{r-k}.
\end{split}
\end{equation}
Define a new set of coordinates by
\begin{equation}
\label{eq23}
\Xi_p=\sum _k Q_kZ_{k-p}.
\end{equation}
Combining eq(\ref{eq21}), eq(\ref{eq22}) and eq(\ref{eq23}), we obtain the equation for the coupled phonons:
\begin{equation}
\label{eq24}
\ddot{\Xi}_p+\sum_{k}\omega_k^2Z_{k-p}\Xi_k=0.
\end{equation}
Since the wave vector $k$ and $p$ are in the Brillouin Zone of the infinitely long chain, the eq.(\ref{eq24}) can be transformed in integral form leading to eq.(\ref{eq5}) in the main text.

\end{document}